\documentclass[10pt]{iopart}
 \usepackage{graphicx}
\usepackage{graphicx,color}

\expandafter\let\csname equation*\endcsname\relax
  \expandafter\let\csname endequation*\endcsname\relax
\usepackage{amsmath}
\usepackage{color}

\usepackage{amssymb}
\usepackage{amsfonts}
\usepackage{bm}
\usepackage[T1]{fontenc}
\usepackage[utf8]{inputenc}
\usepackage{iopams}
\def\<{\langle}
\def\>{\rangle}

\begin{document}

\setlength{\parindent}{0cm}

\title[]{Tuning the collective decay of two entangled emitters by means of a nearby surface}

\author{R. Palacino$^1$, R. Passante$^{1,2}$ and L. Rizzuto$^{1,2}$ }

\address{$^1$ Dipartimento di Fisica e Chimica, Università degli Studi di Palermo, Via Archirafi 36, 90123 Palermo, Italy}

\address{$^2$ INFN, Laboratori Nazionali del Sud, 95123 Catania, Italy}
\author{P. Barcellona$^3$ and S. Y. Buhmann$^{3,4}$}

\address{$^3$ Physikalisches Institut, Albert-Ludwigs-Universit\"{ä}t Freiburg, Hermann-Herder-Str. 3, 79104 Freiburg, Germany}

\address{$^4$ Freiburg Institute for Advanced Studies, Albert-Ludwigs-Universit\"{ä}t Freiburg, Albertstr. 19, 79104 Freiburg, Germany}

\begin{abstract}

We consider the radiative properties of a system of two identical correlated atoms interacting with the electromagnetic field in its vacuum state in the presence of a generic dielectric environment. We suppose that the two emitters are prepared in a symmetric or antisymmetric superposition of one ground state and one excited state and we evaluate the transition rate to the collective ground state, showing distinctive cooperative radiative features. Using a macroscopic quantum electrodynamics approach to describe the electromagnetic field, we first obtain an analytical expression for the decay rate of the two entangled two-level atoms in terms of the Green's tensor of the generic external environment. We then investigate the emission process when both atoms are in free space and subsequently when a perfectly reflecting mirror is present, showing how the boundary affects the physical features of the superradiant and subradiant emission by the two coupled emitters. The possibility to control and tailor radiative processes is also discussed.

\end{abstract}
\submitto{\jpb}

\maketitle
\ioptwocol

\medskip

\section{Introduction}

Spontaneous emission processes by multi-atom systems coherently coupled to the electromagnetic field have been extensively explored in the literature since the seminal work by Dicke in 1954 \cite{dicke}, in which the collective emission of $N$ identical atoms with interatomic separation much shorter than the wavelength of the atomic transition was considered. It was shown that the resulting radiation intensity was proportional to $N^{2}$, rather than $N$ (as expected for the intensity radiated by independent atoms), with a decay time proportional to the inverse of the number of emitters \cite{gross,peng,persico,benedict}. This phenomenon is commonly known as \textit{Superradiance} and it occurs when the sample is prepared in a symmetric superposition of atomic states. The underlying physical process responsible for such enhanced radiative behavior is the constructive interference between emitted waves and researchers have pursued and explored it by studying quantum emitters coupled to various environments including microcavities \cite{temnov,pan} or plasmonic waveguides \cite{martin-cano,fleury} and left-handed metamaterials \cite{kastel} as well as classical emitters near a metal interface \cite{choquette}, with  the potential applications in quantum communication \cite{casabone,chaneliere,jen}.

A closely related process is the collective Lamb Shift of an ensemble of many identical two-level atoms interacting cooperatively with a resonant radiation field. In this case, a virtual photon emitted by one atom may be reabsorbed by another one within the ensemble and, as for the superradiant emission, the process sensitively depends on the spatial arrangement of the system. Contrarily to superradiance, this process is difficult to observe at high atomic densities because of atom-atom interactions which tend to mask it. However, it becomes experimentally accessible in the case of an extended sample, for which $R\gg\lambda_{0}$ ($R$ being the interatomic separation and $\lambda_{0}$ the atomic transition wavelength) as it has been explored for an ensemble of Fe atoms embedded in a low-Q planar cavity \cite{ralf}, for a confined vapor nanolayer of Rb atoms \cite{keaveney} and for a cold-atom ensemble \cite{pellegrino}.

The counterpart of superradiance is \textit{Subradiance} \cite{dicke,gross,crubellier,devoe}, occurring when the system is prepared in an antisymmetric state. In this situation, destructive interference between the neighboring radiators leads to a drastic suppression of the emission intensity, which is vanishing when the interatomic distance is much smaller than the atomic transition wavelength. Contrarily to superradiance, subradiance is more elusive, mainly because the corresponding states are weakly coupled to the environment. They are not affected by decoherence, therefore appealing for quantum information processing. Indeed in Ref. \cite{petrosyan} the authors propose a combined optical/solid-state approach to realize a quantum processor based on "subradiant dimers" of quantum dots, resonantly coupled by dipole-dipole interaction and implanted in low-temperature solid host materials at controllable nanoscale separations. Other works focused on the controlled production of superradiant and subradiant states when the artificial atomic systems are tightly confined in optical lattices \cite{takasu} or in quantum electrodynamics (QED) circuits \cite{filipp}.

In this framework, the main aim of the present paper is to investigate the influence of a structured environment on the collective spontaneous emission of a system of two identical correlated atoms. By using macroscopic quantum electrodynamics \cite{buhmann,knoll,scheel} to characterize the electromagnetic field in the presence of macroscopic magnetodielectric objects, the first result we have obtained is an analytical expression for the decay rate of the joint two atom-field system, prepared in their symmetric (antisymmetric) entangled state in terms of the Green's tensor of a general structured environment.

We then use this result to analyze the radiative behavior of the two entangled quantum emitters in two specific cases: when they are located in free space for any interatomic distance, so recovering the known superradiant and subradiant features in particular in the subwavelength spatial region; when a perfectly reflecting planar surface is present, showing how its presence significantly affects the physical features of the collective spontaneous emission.
We explicitly consider the cases when both atomic dipoles are parallel or perpendicular to the line connecting them.
Our results suggest the possibility to tune and manipulate the system's superradiant and subradiant behavior by suitably placing the two radiators with respect to the reflecting boundary.

The paper is organized as follows. Section 2 outlines the basic theory of macroscopic quantum electrodynamics, presenting the main properties of the Green's tensor we have exploited in our analysis. Section 3 is dedicated to the methodology and results. Through a suitably adapted version of the Wigner-Weisskopf theory \cite{santra} we find the decay probability of the initial state of our joint two atom-field system that we then apply in two specific situations: atoms in free space and in the presence of a perfectly reflecting mirror, illustrated in Sections 4 and 5, respectively. Our concluding remarks are given in Section 6.

\section{Medium-assisted field formalism}

As mentioned, the technique we exploit to describe the body-assisted electromagnetic field and to investigate the collective radiative behavior of our two-atom system embedded in a general macroscopic environment, is based on the Green's tensor formalism \cite{buhmann}. This method has been widely applied in various contexts, ranging from QED to quantum optics since the Green's tensor includes all the properties of the electromagnetic field with which the system interacts. It is thus a general method to study matter-field interaction in any external environment. As an example, this formalism has been recently applied to the study of the dynamical Casimir-Polder force between a chiral molecule and a plate \cite{pablo} and the van der Waals interaction between a ground and an excited atom in a generic environment \cite{pablo1}. In addition, the macroscopic QED approach has been exploited to unveil cooperative effects in a two-atoms system in the presence of left-handed media \cite{kastel} and in determining how the Casimir-Polder interaction energy of a polarizable atom is affected by a metallic surface \cite{haakh}.

In this section, we introduce some properties of the electromagnetic Green's tensor involved, starting with the quantized body-assisted electric field \cite{buhmann,knoll}

\begin{equation}\begin{split}
\mathbf{E}\mathrm{(\mathbf{r})}&=\int_{0}^{\infty}\mathrm{d}\omega\mathbf{E}(\mathbf{r},\omega)+\mathrm{h.c.}=\\
&=\int_{0}^{\infty}d\omega\sum_{\lambda=e,m}\int d^{3}r'\mathbf{G}_{\lambda}(\mathbf{r},\mathbf{r'},\omega)\cdot\mathbf{f}_{\lambda}(\mathbf{r'},\omega)	\\	
&+\mathrm{h.c.}
\label{Equant}
\end{split}\end{equation}

where $\mathbf{G}(\mathbf{r},\mathbf{r^{\prime}},\omega)$ is the Green's tensor, $\mathbf{f}_{\lambda}(\mathbf{r},\omega)$ and $\mathbf{f}_{\lambda}^{\dagger}(\mathbf{r},\omega)$ are bosonic annihilation and creation operators representing the collective excitations of the body-field system, satisfying the following commutation rules

\begin{equation}
\left[\mathbf{f}_{\lambda}(\mathbf{r},\omega),\mathbf{f}_{\lambda'}(\mathbf{r'},\omega')\right]=0,
\label{commrul}
\end{equation}

\begin{equation}
\left[\mathbf{f}_{\lambda}(\mathbf{r},\omega),\mathbf{f}_{\lambda'}^{\text{\dag}}(\mathbf{r'},\omega')\right]=\delta_{\lambda\lambda'}\bm{\delta}(\mathbf{r\mathrm{-}r')\delta(\omega-\omega')}.
\label{complconiug}
\end{equation}

where the subscript $\lambda=e,m$ identifies the electric and magnetic parts.
The $\mathbf{G}_{\lambda}(\mathbf{r},\mathbf{r'},\omega)$ obeys the following integral relation

\begin{equation}\begin{split}
\sum_{\lambda=e,m}\int d^{3}s\mathbf{G}_{\lambda}(\mathbf{r},\mathbf{s},\omega)&\cdot\mathbf{G}_{\lambda}^{*T}(\mathbf{r^{\prime}},\mathbf{s},\omega)=\\
&=\frac{\hbar\mu_{0}}{\pi}\omega^{2} \textrm{Im}\mathbf{G}(\mathbf{r},\mathbf{r^{\prime}},\omega).
\label{intrel}
\end{split}\end{equation}

According to the geometry of the chosen environment, generally one needs to consider two contributions to the total tensor $\mathbf{G}(\mathbf{r},\mathbf{r^{\prime}},\omega)$: $\mathbf{G}^{(0)}(\mathbf{r},\mathbf{r^{\prime}},\omega)$, describing the electromagnetic field in the bulk region of the medium, and a scattering part $\mathbf{G}^{(1)}(\mathbf{r},\mathbf{r^{\prime}},\omega)$ accounting for additional contributions due to reflections at or transmission through the boundaries. For the purpose of this work, a perfectly reflecting planar surface will be considered as macroscopic environment for the two emitters. The analytical expressions for the bulk and scattering parts of the Green's tensor are, respectively, the free space Green's tensor \cite{buhmann}, \cite{knoll}

\begin{equation}\begin{split}
\mathbf{G}^{(0)}(\mathbf{r},\mathbf{r^{\prime}},\omega)=&-\frac{c^{2}}{3\omega^{2}}\boldsymbol{\delta}(\boldsymbol{\varrho})-\frac{c^{2}e^{i\omega\varrho/c}}{4\pi\omega^{2}\varrho^{3}}\left\{ \left[1-i\frac{\omega\varrho}{c}\right.\right.\\
&\hspace{-1cm}\left.\left.-\left(\frac{\omega\varrho}{c}\right)^{2}\right]\mathbf{I}-\left[3-3i\frac{\omega\varrho}{c}-\left(\frac{\omega\varrho}{c}\right)^{2}\right]\mathbf{e}_{\varrho}\mathbf{e}_{\varrho}\right\}
\label{Gvacuum}
\end{split}\end{equation}

where $\boldsymbol{\varrho}=\mathbf{r}-\mathbf{r}^{\prime}$, $\varrho=\left|\boldsymbol{\varrho}\right|$, $\mathbf{e}_{\varrho}=\frac{\boldsymbol{\varrho}}{\left|\boldsymbol{\varrho}\right|}$ and \cite{buhmann}

\begin{equation}\begin{split}
\mathbf{G}^{(1)}(\mathbf{r},\mathbf{r^{\prime}},\omega)=\mathbf{G}^{(0)}(\mathbf{r},\mathbf{r^{\prime*}},\omega)\cdot \mathbf{R},
\label{G1}
\end{split}\end{equation}

with $\mathbf{r^{\prime*}}$ being the position of the source's image behind the mirror and $\mathbf{R}$ the reflection matrix defined by

\begin{equation}
\mathbf{R}=\left(\begin{array}{ccc}
-1 & 0 & 0\\
0 & -1 & 0\\
0 & 0 & 1
\end{array}\right).
\end{equation}

The scattering component $\mathbf{G}^{(1)}(\mathbf{r},\mathbf{\mathbf{r^{\prime}}},\omega)$ is related to the free-space Green's tensor at the distance $\varrho_{+}$, that is the distance that a reflected wave on the mirror needs to travel from one atom to reach the other. An alternative interpretation is that the interaction occurs between one atom, say A, located at $\mathbf{r}$, and the mirror image of atom B, located at $\mathbf{r^{\prime*}}$, behind the plate.

Moreover, given two oriented dipoles $\mathbf{d}_{A}=(d_{Ax},d_{Ay},d_{Az})$ and $\mathbf{d}_{B}=(d_{Bx},d_{By},d_{Bz})$, their images are constructed by a spatial reflection at the $z=0$ plane, together with an interchange of positive and negative charges:

\begin{equation}
\mathbf{d}_{A}^{*}=\mathbf{R}\cdot\mathbf{d}_{A},
\label{dA*}
\end{equation}

\begin{equation}
\mathbf{d}_{B}^{*}=\mathbf{R}\cdot\mathbf{d}_{B}.
\label{dB*}
\end{equation}

Two additional useful properties are the following:
the free space Green's tensor is symmetric under exchange of its spatial arguments

\begin{equation}
\mathbf{G}^{(0)}(\mathbf{r},\mathbf{r^{\prime}},\omega)=\mathbf{G}^{(0)T}(\mathbf{r^{\prime}},\mathbf{r},\omega)=\mathbf{G}^{(0)}(\mathbf{r^{\prime}},\mathbf{r},\omega),
\label{symmetric}
\end{equation}

a key feature in our computation of the system's collective decay rate in the following; its imaginary part is finite for $\mathbf{r}=\mathbf{r}^{\prime}$ and reads

\begin{equation}
\textrm{Im}\mathbf{G}^{(0)}(\mathbf{r},\mathbf{r},\omega)=\frac{\omega}{6\pi c}\mathbf{I}.
\label{ImGvacuum}
\end{equation}

with $\mathbf{I}$ being the unit tensor.

The next sections will focus on the methodology we use to evaluate the collective decay rate of the two entangled atoms when they are located in an arbitrary macroscopic environment, first, and subsequently when a perfectly reflecting surface is present.

\section{The collective spontaneous decay rate}

We use an approach based on second-order time-dependent perturbation theory holding for a general physical system in the presence of an external perturbation. It leads to the following result

\begin{equation}
C_{i}(t)=C_{i}(0)e^{-i(\triangle_{i}-i\varGamma_{i}/2)t/\hbar}
\label{c(i)}
\end{equation}

for the probability $C_{i}(t)$ to find the system in its initial state $\left|i\right\rangle$, of energy $\hbar\omega_{0}$, at some instant of time $t$ \cite{santra}. The above expression contains an energy shift $\triangle_{i}$ and the decay rate of the amplitude $C_{i}(t)$, given by

\begin{equation}
\varDelta_{i}=\textrm{P.V.}\left(\sum_{I\neq i}\frac{\left|\left\langle I\right|\hat{H}_{1}\left|i\right\rangle \right|^{2}}{\hbar(\omega_{i}-\omega_{I})}\right)
\label{delta}
\end{equation}

\begin{equation}
\varGamma_{i}=\frac{2\pi}{\hbar}\sum_{I\neq i}\left|\left\langle I\right|\hat{H}_{1}\left|i\right\rangle \right|^{2}\delta(\hbar[\omega_{i}-\omega_{I}]).
\label{gamma}
\end{equation}

respectively, where $\hat{H}_{1}$ is the perturbation, $\textrm{P.V.}$ indicates the principal value and $\left|I\right\rangle$ the intermediate states involved in the process. In the following we will focus on the decay rate.

We thus first define the total Hamiltonian and the initial state vectors of our composite two-atoms and field system

\begin{equation}
\hat{H}=\hat{H}_{0}+\hat{H}_{1}=\sum_{\xi=A,B}\hat{H}_{at}(\xi)+\hat{H}_{F}+\sum_{\xi=A,B}\hat{H}_{int}(\xi)
\label{Hamilt.}
\end{equation}

where $\hat{H}_{at}(\xi)$ is the Hamiltonian of the atomic species $\xi$, $\hat{H}_{F}$ the field Hamiltonian and $\hat{H}_{1}$ the interaction Hamiltonian of the two atoms with the electromagnetic field.
We consider electric atoms and adopt the multipolar coupling scheme in the dipole approximation, so that

\begin{equation}
\hat{H}_{1}=-\mathbf{d}_{A}\cdot\mathbf{E}(\mathbf{r}_{A})-\mathbf{d}_{B}\cdot\mathbf{E}(\mathbf{r}_{B})
\label{dipolint}
\end{equation}

with $\mathbf{d}_{\xi}$ the transition dipole moment operator of atom $\xi$ and $\mathbf{E}(\mathbf{r}_{\xi})$ the electric field operator evaluated at the atomic position $\mathbf{r}_{\xi}$.
In addition, we need to specify the initial state of the system. A generic eigenvector of $\hat{H}_{0}$ is expressed as product of atomic and field states with eigenenergy given by the sum of the energy of the atomic species and of the photons in some field modes. Here, we consider two identical atoms prepared in the correlated symmetric (antisymmetric) state

\begin{equation}
\left|i_{\pm}\right\rangle =\frac{1}{\sqrt{2}}\left(\left|e_{A},g_{B};\left\{ 0\right\} \right\rangle \pm\left|g_{A},e_{B};\left\{ 0\right\} \right\rangle \right)
\label{statosimm}
\end{equation}

with the atomic excitation being delocalized over the two atoms, and the field in the vacuum state ($e$ and $g$ indicate ground and excited states, respectively). These are the superradiant (symmetric) and subradiant (antisymmetric) states in the Dicke model.

Due to the expression of the interaction Hamiltonian, only two intermediate states contribute

\begin{equation}
\left|I_{1}\right\rangle =\left|g_{A},g_{B};1_{\lambda}(\mathbf{r},\omega)\right\rangle, \quad \left|I_{2}\right\rangle =\left|e_{A},e_{B};1_{\lambda}(\mathbf{r},\omega)\right\rangle
\label{interm12}
\end{equation}

so that, using the expression of the electric field in terms of the Green's tensor, given in equation (\ref{Equant}), the matrix elements for the intermediate states $\left|I_{1}\right\rangle$ and $\left|I_{2}\right\rangle$ can be now evaluated. By collecting equations (\ref{Equant}),(\ref{dipolint}),(\ref{statosimm}) and (\ref{interm12}), one finds the matrix element for the symmetric correlated state

\begin{equation}\begin{split}
\left\langle i_+\right|\hat{H}_{1}\left|I_{1}\right\rangle =&-\frac{1}{\sqrt{2}}\Big[\mathbf{d}_{\mathrm{A}}^{\mathrm{eg}}\cdot\mathbf{G}_{\lambda}(\mathbf{r}_{\mathrm{A}},\mathbf{r},\omega)\\
&+\mathrm{\mathbf{d}_{\mathrm{B}}^{\mathrm{eg}}\cdot\mathbf{G}_{\lambda}(\mathbf{r}_{\mathrm{B}},\mathbf{r},\omega)}\Big]
\label{element1new}
\end{split}\end{equation}

where we have assumed that the dipole operator $\mathbf{d}$ has only off-diagonal matrix elements due to selection rules and $\mathbf{d_{\xi}}^{\mathrm{eg}}$ are matrix elements between the excited and the ground state.

To compute $\left\langle I_{1}\right|\hat{H}_{1}\left|i_{+}\right\rangle $ it is sufficient to take the hermitian conjugate of (\ref{element1new}).
Similarly, the matrix element containing the second intermediate state $\left|I_{2}\right\rangle$ is

\begin{equation}\begin{split}
\left\langle i_+\right|\hat{H}_{1}\left|I_{2}\right\rangle =&-\frac{1}{\sqrt{2}}\Big[\mathbf{d}_{\mathrm{B}}^{\mathrm{ge}}\cdot\mathbf{G_{\lambda}(\mathbf{r}_{\mathrm{B}},}\mathbf{r},\omega)\\
&+\mathrm{\mathbf{d}_{\mathrm{A}}^{\mathrm{ge}}\cdot\mathbf{G_{\lambda}(\mathbf{r}_{\mathrm{A}},}\mathbf{r},\omega)}\Big]
\label{element2}
\end{split}\end{equation}

and the hermitian conjugate can be computed. The matrix elements for the antisymmetric correlated state can be found in an analogous way.

By exploiting the integral relation fulfilled by the Green's tensor equation (\ref{intrel}), from (\ref{gamma}) we can evaluate the decay rate of the initial state of the system, when the two atoms are placed in the environment described by the Green's tensor $\mathbf{G}(\mathbf{r},\mathbf{r'},\omega)$ \cite{kastel}:

\begin{equation}\begin{split}
\varGamma_{i\pm}=&\frac{\mu_{0}\omega_{0}^{2}}{\hbar}\Big[\mathbf{d}_{A}^{eg}\cdot \textrm{Im}\mathbf{G}(\mathbf{r}_{A},\mathbf{r}_{A},\omega_{0})\cdot\mathbf{d}_{A}^{ge}\\
&+\mathbf{d}_{B}^{eg}\cdot \textrm{Im}\mathbf{G}(\mathbf{r}_{B},\mathbf{r}_{B},\omega_{0})\cdot\mathbf{d}_{B}^{ge}\\
&\pm\mathbf{d}_{A}^{eg}\cdot \textrm{Im}\mathbf{G}(\mathbf{r}_{A},\mathbf{r}_{B},\omega_{0})\cdot\mathbf{d}_{B}^{ge}\\
&\pm\mathbf{d}_{B}^{eg}\cdot \textrm{Im}\mathbf{G}(\mathbf{r}_{B},\mathbf{r}_{A},\omega_{0})\cdot\mathbf{d}_{A}^{ge}\Big]
\label{decaycompact form simm}
\end{split}\end{equation}

with $\omega_{0}$ being the atomic transition frequency and the $+$ or $-$ signs refer to the symmetric or antisymmetric state.

When the atoms satisfy time-reversal symmetry $\mathbf{d}^{ge}=\mathbf{d}^{eg}$, the above expression (\ref{decaycompact form simm}) can be written in the following form

\begin{equation}
\varGamma_{i\pm}=\frac{\varGamma_{A}+\varGamma_{B}}{2}\pm \varGamma_{AB}.
\label{generalform3}
\end{equation}
where:
\begin{equation}\varGamma_{A}=\frac{2\mu_{0}\omega_{0}^{2}}{\hbar}\mathbf{d}_{A}^{eg}\cdot \textrm{Im}\mathbf{G}(\mathbf{r}_{A},\mathbf{r}_{A},\omega_{0})\cdot\mathbf{d}_{A}^{ge}\end{equation}
is the single-atom decay rate and
\begin{equation}\varGamma_{AB}=\frac{2\mu_{0}\omega_{0}^{2}}{\hbar}\mathbf{d}_{A}^{eg}\cdot \textrm{Im}\mathbf{G}(\mathbf{r}_{A},\mathbf{r}_{B},\omega_{0})\cdot\mathbf{d}_{B}^{ge}\end{equation}
is an interference term.

Equations (\ref{decaycompact form simm})  describe the overall decay rates of the two atom-field system prepared in a symmetric and antisymmetric state, where the excitation energy is delocalized over the two atoms. These relations are given in terms of the Green's tensor of a generic structured environment surrounding the system and contain both terms evaluated at the position of the single atoms, corresponding to the decay rates of independent atoms, and terms describing interference effects, as we shall discuss in detail in the following sections.
The reason for the non-vanishing interference term is that, in the transition from the correlated initial state to the intermediate state, it is not possible to know which atom emits the photon.

\section{Atoms in free space}

The method we followed in the previous paragraph has led to the expression (\ref{generalform3}) which can be applied to any external environment, provided the form of the Green's tensor of the specific environment is known.

We first analyze the behaviour of the decay rate in free space varying the interatomic separation and hence observing it in the non-retarded short-distance limit ($\varrho=\left|\mathbf{r}_{B}-\mathbf{r}_{A}\right|\ll c/\omega_{0}$)  and in the retarded long-distance limit ($\varrho=\left|\mathbf{r}_{B}-\mathbf{r}_{A}\right|\gg c/\omega_{0}$). This allows us to recover the known superradiant and subradiant behaviours.

We use the free-space Green's tensor $\mathbf{G}(\mathbf{r},\mathbf{r'},\omega)=\mathbf{G}^{(0)}(\mathbf{r},\mathbf{r'},\omega)$, choosing the axis in order to have $\mathbf{r_{\mathrm{A}}=\mathrm{(0,0,z_{A})}}$ and $\mathbf{r_{\mathrm{B}}=\mathrm{(0,0,z_{B})}}$, then $\boldsymbol{\varrho}=\mathbf{r}_{A}-\mathbf{r}_{B}=(0,0,z)$ with $z=z_{A}-z_{B}$.
Furthermore, we consider dipole moments with arbitrary spatial orientations: $\mathbf{d}_{A}=(d_{Ax},d_{Ay},d_{Az})$ and $\mathbf{d}_{B}=(d_{Bx},d_{By},d_{Bz})$.

Since the free space Green's tensor is symmetric with respect to exchange of its position arguments, the mixed terms in equation (\ref{decaycompact form simm}) coincide, thus giving the collective decay rate in the following form

\begin{equation}\begin{split}
\varGamma_{i+}=& \frac{\mu_{0}\omega_{0}^{2}}{\hbar}\Big[\mathbf{d}_{A}^{eg}\cdot \textrm{Im}\mathbf{G}^{0)}(\mathbf{r}_{A},\mathbf{r}_{A},\omega_{0})\cdot\mathbf{d}_{A}^{ge}\\
&+\mathbf{d}_{B}^{eg}\cdot \textrm{Im}\mathbf{G}^{0)}(\mathbf{r}_{B},\mathbf{r}_{B},\omega_{0})\cdot\mathbf{d}_{B}^{ge}\\
&+2\mathbf{d}_{A}^{eg}\cdot \textrm{Im}\mathbf{G}^{0)}(\mathbf{r}_{A},\mathbf{r}_{B},\omega_{0})\cdot\mathbf{d}_{B}^{ge}\Big]
\label{compactform2}
\end{split}\end{equation}

Equivalently

\begin{equation}
\varGamma_{i+}=\frac{\varGamma_{A}+\varGamma_{B}}{2}+\varGamma_{AB},
\label{compactform3}
\end{equation}

by identifying the single-atom contributions $\varGamma_{A}$ and $\varGamma_{B}$ (free-space spontaneous decay of an independent excited atom), and the "interference" term $2\varGamma_{AB}$ by the equations:

\begin{align}\nonumber
\varGamma_{A}=& \frac{\left| \mathbf{d}_A^{eg} \right|^2\omega _0^3}{3\pi \varepsilon _0\hbar c^3},\\\nonumber
\varGamma_{AB} =& \frac{1}{2\pi \varepsilon _0\hbar z^3}\Big( \left( \mathbf{d}_A^{eg} \cdot \mathbf{d}_B^{ge} - 3d_{Az}^{eg} d_{Bz}^{ge} \right) \\ \nonumber
 \times \big( \lambda\cos \lambda &- \sin \lambda \big)+ \left( \mathbf{d}_A^{eg} \cdot \mathbf{d}_B^{ge} - d_{Az}^{eg} d_{Bz}^{ge} \right)\lambda^2\sin \lambda\Big)\\
\end{align}
where $\lambda=z \omega/c$ \cite{tanas}.

Firstly, we take the imaginary part of the free space Green's tensor (\ref{ImGvacuum}) \cite{buhmann} which will hence be applied to the first two terms on the right-hand side of equation (\ref{compactform2}). Secondly, we consider the two asymptotic behaviors, starting with the nonretarded limit $\varrho\ll c/\omega_{0}$.  When $\mathbf{r}_{A}\rightarrow\mathbf{r}_{B}$ (short-distance regime), $\textrm{Im}\mathbf{G}(\mathbf{r}_{A},\mathbf{r}_{B},\omega_{0})\rightarrow\frac{\omega_0}{6\pi c}\mathbf{I}$ and the interference term in equation (\ref{compactform2}) yields $\varGamma_{AB}\rightarrow \frac{\varGamma_{A}+\varGamma_{B}}{2}$ (when $\mathbf{d}_{A}$ and $\mathbf{d}_{B}$ are parallel, otherwise it vanishes) so that

\begin{equation}
\varGamma_{i+}^{\text{nret}}=\varGamma_{A}+\varGamma_{B}.
\label{double}
\end{equation}

In the non-retarded limit of very small interatomic distances, the transition rate of the system to the collective ground state is the sum of the two individual atom's rates, showing the superradiant decay of the initial state (\ref{statosimm}).

In the opposite case, the retarded limit $\varrho\gg c/\omega_{0}$, the interference term vanishes due to the Green's tensor boundary condition $\mathbf{G}^{(0)}(\mathbf{r},\mathbf{r'},\omega)\rightarrow0$ for $\left|\mathbf{r}-\mathbf{\mathbf{r'}}\right|\rightarrow\infty$. Therefore, we find

\begin{equation}
\varGamma_{i+}^{\text{ret}}= \frac{\varGamma_{A}+\varGamma_{B}}{2}
\label{single}
\end{equation}

\begin{figure}
\begin{centering}
  \includegraphics[scale=0.45]{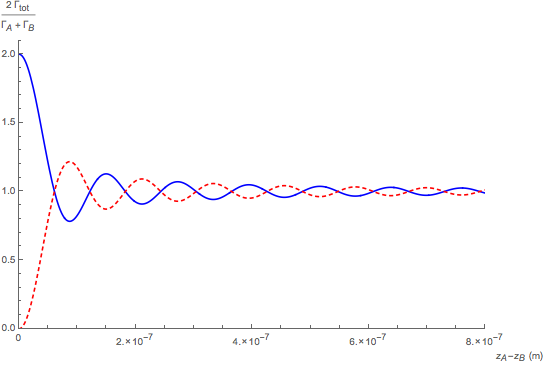}
  \par\end{centering}
  \caption{Comparison between the transition rates of the symmetric (blue continuous line) and antisymmetric states (red dashed line) in free space, showing respectively in the non-retarded limit, the superradiant and subradiant behaviours for two aligned atomic dipole moments.}
\label{confrontovuoto}
\end{figure}

For intermediate distances $\varrho$, the decay rate displays an oscillatory behaviour due to the presence of periodic functions in the imaginary part of the free-space Green's tensor, scaling with the interatomic distance $z$ as $\frac{\sin\left(\frac{z\omega_{0}}{c}\right)}{z}$. The oscillation will be damped as the two atoms are further far apart, yielding the independent-atoms decay at large distances. In Figure \ref{confrontovuoto}, we display the ratio between the total decay rate and the sum of decay rates from independent atoms, so that the non-retarded and retarded limit behaviours are clearly visible. We considered the transition frequency for the $2p\rightarrow1s$ transition of the Hydrogen atom.

With regard to the blue continous curve, a peak is observed for the minimum atomic separation considered (see equation (\ref{double})). This is a signature of the cooperative behaviour (superradiance) due to a constructive correlation between the two identical atoms. The emission rate rapidly decreases with increasing distances, leading to the ordinary spontaneous emission of independent atoms in the limit of large interatomic separations.

Furthermore, whenever the distance between the two atoms is $\varrho=\left(n+1/4\right)\lambda_{0}$ with $n\in\mathbb{N}^{+}$ positive integer, the transition rate is increased, compared to that of independent atoms. On the contrary, when $\varrho=\left(m+3/4\right)\lambda_{0}$ with $m\in\mathbb{N}$, we observe a reduction of the transition probability rate. The points corresponding to $\varrho=n\lambda_{0}/2$ indicate a vanishing interference term. Analogue behaviours are observed in \cite{arias} where the authors study the emission rate of a bi-atomic system prepared in a correlated symmetric state and interacting with the massless scalar field.

The transition rate from the antisymmetric state to the collective ground state in free space is given by

\begin{equation}\begin{split}
\varGamma_{i-}=&\frac{\mu_{0}\omega_{0}^{2}}{\hbar}\Big[\mathbf{d}_{A}^{eg}\cdot \textrm{Im}\mathbf{G}(\mathbf{r}_{A},\mathbf{r}_{A},\omega_{0})\cdot\mathbf{d}_{A}^{ge}\\
&+\mathbf{d}_{B}^{eg}\cdot \textrm{Im}\mathbf{G}(\mathbf{r}_{B},\mathbf{r}_{B},\omega_{0})\cdot\mathbf{d}_{B}^{ge}\\
&-2\mathbf{d}_{A}^{eg}\cdot \textrm{Im}\mathbf{G}(\mathbf{r}_{A},\mathbf{r}_{B},\omega_{0})\cdot\mathbf{d}_{B}^{ge}\Big]
\label{compactformantisimm}
\end{split}\end{equation}

that we can write as

\begin{equation}
\varGamma_{i-}=\frac{\varGamma_{A}+\varGamma_{B}}{2}-\varGamma_{AB}.
\label{compactformantisimm1}
\end{equation}

In this case, the behaviour is completely different for the non-retarded limit $\varrho\ll c/\omega_{0}$. In fact, since in this limit $2\varGamma_{AB}\rightarrow\varGamma_{A}+\varGamma_{B}$, a complete inhibition of the spontaneous transition rate due to destructive interference of quantum correlations between the two atoms is observed, recovering the known subradiant behaviour. In contrast with the previous case, we get a lower transition rate for $\varrho=\left(n+1/4\right)\lambda_{0}$, while for $\varrho=\left(m+3/4\right)\lambda_{0}$ the interference effects lead to an enhancement of the output rate, as displayed in Figure \ref{confrontovuoto} (red dashed line). As before, the points corresponding to $\varrho=n\lambda_{0}/2$ indicate a vanishing interference term. Similar results for the antisymmetric state are also obtained in the massless scalar field case discussed in \cite{arias}.

In both cases we find that the quantum interference term between the atoms produces vanishing contributions for large interatomic separations $\varrho\gg c/\omega_{0}$. The influence of the quantum interference is stronger for short distances between the atoms, compared to their transition wavelength.
In the free-space case, our results thus agree with those reported in Refs.\cite{dicke,gross} with different methods. Superradiance has been observed in experiments both with atoms \cite{skribanowitz} and quantum dots \cite{Scheibner}.

\section{Atoms near a perfectly reflecting plate}

Having obtained in Section 3 the general expression for the decay rate of the two entangled atoms in terms of the Green's tensor, we can consider other environments by using the relative expression of $\mathbf{G}(\mathbf{r},\mathbf{r^{\prime}},\omega)$. In this section we will focus on how the cooperative emission rate is affected when a perfectly reflecting mirror is present.

\begin{figure}
\begin{centering}
  \includegraphics[scale=0.55]{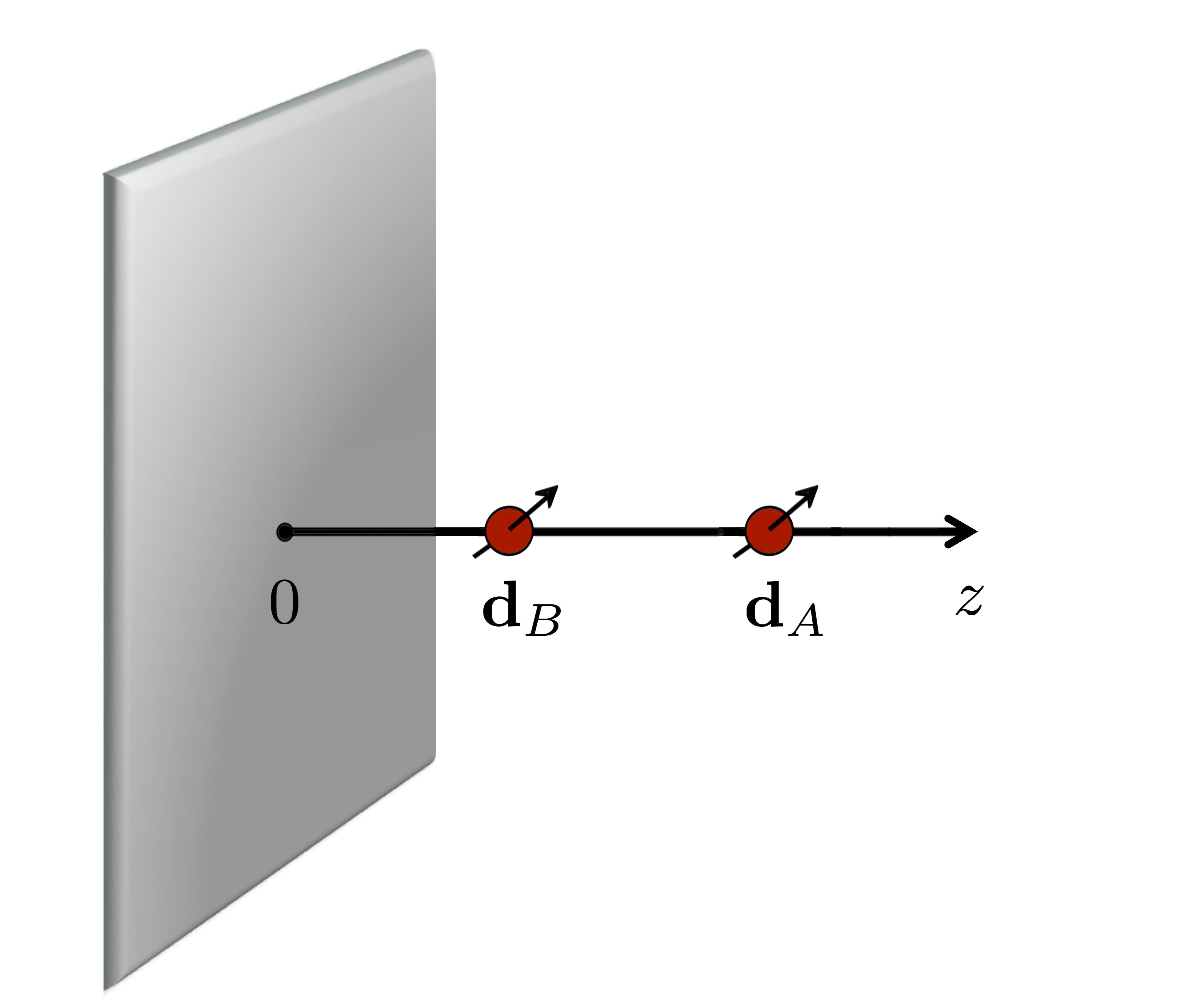}
  \par\end{centering}
  \caption{Emitting dipoles of the two entangled identical atoms in the presence of a perfectly reflecting mirror.}
\label{imagerob}
\end{figure}

Let us consider both atoms aligned along the $z$-axis perpendicular to the surface as shown in Figure \ref{imagerob}. In addition to the interatomic distance, the presence of the mirror introduces a new spatial parameter: $\varrho_{+}=z_{A}+z_{B}$, that is the distance between one atom and the mirror image of the other behind the plate.

In this physical configuration, two terms contribute to the system's decay rate

\begin{equation}
\Gamma=\Gamma^{(0)}+\Gamma^{(1)},
\label{Gamma}
\end{equation}

the first one on the right hand side is due to the free-space interaction between the atoms and the second to the presence of the boundary, which importantly modifies the modes of the electromagnetic field.
Each term in equation (\ref{Gamma}) refers to the respective component of the total Green's tensor

\begin{equation}
\mathbf{G}(\mathbf{r},\mathbf{r^{\prime}},\omega)=\mathbf{G}^{(0)}(\mathbf{r},\mathbf{r^{\prime}},\omega)+\mathbf{G}^{(1)}(\mathbf{r},\mathbf{r^{\prime}},\omega),
\label{Gtot}
\end{equation}

where the scattering component $\mathbf{G}^{(1)}(\mathbf{r},\mathbf{r^{\prime}},\omega)$ is given in equation (\ref{G1}).

Taking into account equation (\ref{G1}) the modification of the decay due to the presence of the mirror can be thought of as an effective cooperative emission from the correlated state of one atom and the mirror image of the other atom.

By recalling equation (\ref{decaycompact form simm}) (the same procedure can be used also for the decay rate from the antisymmetric state), and taking the imaginary part of $\mathbf{G}(\mathbf{r},\mathbf{r^{\prime}},\omega)$, we consider the non-retarded ($r_{A},r_{B},\varrho ,\varrho_{+}\ll c/\omega_{0}$) and the retarded regime ($r_{A},r_{B},\varrho ,\varrho_{+}\gg c/\omega_{0}$), evaluating first the Green's tensor in these two limits.

Let us start with the non-retarded case. By recalling the property (\ref{ImGvacuum}), holding for $\mathbf{r}=\mathbf{r}^{\prime}$  and observing that for $\mathbf{r}_{A}\rightarrow \mathbf{r}_{A^{*}}$, $\mathbf{r}_{B}\rightarrow \mathbf{r}_{B^{*}}$

\begin{equation}
\textrm{Im}\mathbf{G}^{(1)}(\mathbf{r}_{A},\mathbf{r}_{A},\omega_{0})=\textrm{Im}\mathbf{G}^{(1)}(\mathbf{r}_{B},\mathbf{r}_{B},\omega_{0})=\frac{\omega_{0}}{6\pi c}\mathbf{R}
\label{ImGscatt}
\end{equation}

and for $\mathbf{r}_{A}\rightarrow \mathbf{r}_{B}$ we have also $\mathbf{r}_{A}\rightarrow \mathbf{r}_{B^{*}}$ so that

\begin{equation}
\textrm{Im}\mathbf{G}^{(0)}(\mathbf{r}_{A},\mathbf{r}_{B},\omega_{0})=\frac{\omega_{0}}{6\pi c}\mathbf{I},
\label{ImGbulk}
\end{equation}

\begin{equation}
\textrm{Im}\mathbf{G}^{(1)}(\mathbf{r}_{A},\mathbf{r}_{B},\omega_{0})=\frac{\omega_{0}}{6\pi c}\mathbf{R},
\label{ImGbulk}
\end{equation}

the total decay rate in the non-retarded limit reads

\begin{equation}\begin{split}
\varGamma_{i+}^{\text{nret}}&=\frac{\mu_{0}\omega_{0}^{2}}{\hbar}\left[\mathbf{d}_{A}^{eg}\cdot\left(\frac{\omega_{0}}{6\pi c}\mathbf{I}+\frac{\omega_{0}}{6\pi c}\mathbf{R}\right)\cdot\mathbf{d}_{A}^{ge}\right.\\
&\left.+\mathbf{d}_{B}^{eg}\cdot\left(\frac{\omega_{0}}{6\pi c}\mathbf{I}+\frac{\omega_{0}}{6\pi c}\mathbf{R}\right)\cdot\mathbf{d}_{B}^{ge}\right.\\
&\left.+2\mathbf{d}_{A}^{eg}\cdot\left(\frac{\omega_{0}}{6\pi c}\mathbf{I}+\frac{\omega_{0}}{6\pi c}\mathbf{R}\right)\cdot\mathbf{d}_{B}^{ge}\right]\\
&=\frac{\mu_{0}\omega_{0}^{2}}{\hbar}\frac{\omega_{0}}{6\pi c}\Big[\mathbf{d}_{A}^{eg}\cdot\mathbf{d}_{A}^{ge}+ \mathbf{d}_{A}^{eg}\cdot\mathbf{d}_{A}^{ge*}\\
&+\mathbf{d}_{B}^{eg}\cdot\mathbf{d}_{B}^{ge}+ \mathbf{d}_{B}^{eg}\cdot\mathbf{d}_{B}^{ge*}+2\mathbf{d}_{A}^{eg}\cdot\mathbf{d}_{B}^{ge}\\
&+2\mathbf{d}_{A}^{eg}\cdot\mathbf{d}_{B}^{ge*}\Big],
\end{split}
\label{NR}
\end{equation}

where we have used equations (\ref{dA*},\ref{dB*}).

For what concerns the retarded limit, we consider the following configuration: one atom, say B, in a fixed position close to the surface and $r_{A},\varrho ,\varrho_{+}\gg c/\omega_{0}$. We thus estimate the decay rate to be

\begin{equation}\begin{split}
\varGamma_{i+}^{\text{ret}}&=\frac{\mu_{0}\omega_{0}^{2}}{\hbar}\frac{\omega_{0}}{6\pi c}\Big[\mathbf{d}_{A}^{eg}\cdot\mathbf{I}\cdot
\mathbf{d}_{A}^{ge}+\mathbf{d}_{B}^{eg}\cdot\left(\mathbf{I}+\mathbf{R}\right)\cdot\mathbf{d}_{B}^{ge}\Big]\\
&=\frac{\mu_{0}\omega_{0}^{2}}{\hbar}\frac{\omega_{0}}{6\pi c}\Big[\mathbf{d}_{A}^{eg}\cdot
\mathbf{d}_{A}^{ge}+\mathbf{d}_{B}^{eg}\cdot\mathbf{d}_{B}^{ge}\\
&+\mathbf{d}_{B}^{eg}\cdot\mathbf{d}_{B}^{ge*}\Big].
\label{decayretarded}
\end{split}\end{equation}

Besides the self-interaction due to the free-space component of the Green's tensor, there is an additional term, due to the proximity of atom B to the plate, containing ($\mathbf{d}_{B}^{eg}\cdot\mathbf{d}_{B}^{ge*}$), while atom A presents only the free-space contribution due to its very large distance from the plate.

On the other hand, if atom B is very distant from the surface too, the free-space behaviour of the spontaneous emission rate, obtained in the previous section, is recovered, that is $\Gamma\rightarrow\Gamma^{(0)}$.

\begin{figure}
\begin{centering}
  \includegraphics[scale=0.36]{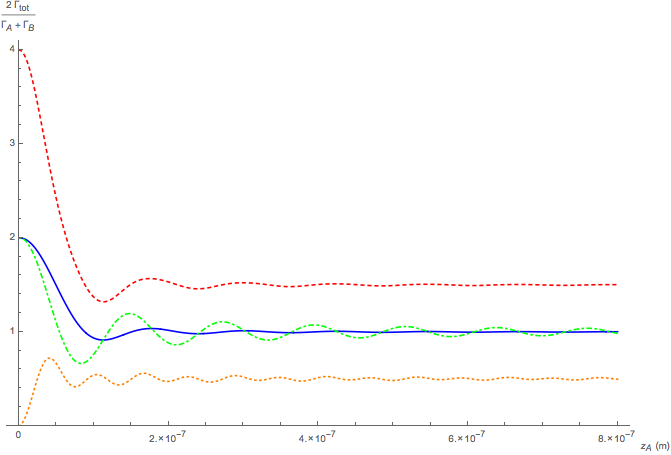}
  \par\end{centering}
\caption{Scaled collective spontaneous emission decay rates (symmetric state), as a function of the position of atom A when atom B is fixed at $z_{B}=10 \mathring{A}$. The red dashed line and the blue continuous line refer, respectively, to the case when the mirror at $z=0$ is present and the free-space case, for dipoles along the $z$-axis (perpendicular to the wall). The orange dotted line and the green dot-dashed line refer, respectively, to the case when the mirror is present and the free-space case, for dipoles along the $x$-axis (parallel to the wall).}

\label{dipolizxsimmetrico}
\end{figure}

Figure \ref{dipolizxsimmetrico} shows a plot of the decay rate for the symmetric state for increasing distance of one of the two atoms (A) with respect to the mirror, when the other atom (B) is at a fixed distance from the plate ($z_{B}=10 \mathring{A}$). The distances of both  atoms from the mirror are compatible with the dipole approximation and the assumption of a perfectly reflecting mirror. Note that in this plot and in the following ones, we assume an atomic transition angular frequency equal to that of the $2p\rightarrow 1s$ transition in the Hydrogen atom
($\omega_0 \simeq 1.55 \cdot 10^{16} \mbox{s}^{-1}$, $\lambda_0 =c/(2\pi \omega_0) \simeq 1.2 \cdot 10^{-7} \mbox{m})$,
a minimum interatomic separation $z_{A}-z_{B}=10 \mathring{A}$, and we consider
perpendicular and parallel orientations of both (identical) atomic dipoles with respect to the line connecting them. Controlling distance and dipole orientation of a quantum emitter with respect to an external environment, for example a nanoparticle, is experimentally achievable and it has been used to obtain enhancement of the spontaneous emission rate \cite{Novotny11}.
As Figure \ref{dipolizxsimmetrico} shows, in the near zone
(all distances much smaller than $\lambda_0$), the decay rate in the case of dipole moments perpendicular to the wall is essentially doubled with respect to the free-space case (red dashed line and blue continuous line, respectively). This can be explained in terms of additional constructive interference between the emitters and their mirror images, since the image dipole of $\mathbf{d}_{\perp}$ coincides with $\mathbf{d}_{\perp}$. If emitter A is far from the boundary while B remains close to it, the interference term $\Gamma_{AB}\simeq 0$; however, the scaled decay rate is greater than the unity value of the empty-space case, as Figure \ref{dipolizxsimmetrico} shows. In fact, in addition to the free-space decay term $\Gamma_{A}^{(0)}$ of atom A, we must consider the term $\Gamma_{B}^{(1)}= \Gamma_{B}^{(0)} $, due to the proximity of atom B to the plate, therefore doubling the decay term related to the emitter B.
A different result is obtained for a parallel alignment configuration, where the overall decay rate is suppressed in the near zone of both atoms (orange dotted line), since the image dipole of $\mathbf{d}_{\Vert}$ is $-\mathbf{d}_{\Vert}$ and therefore their sum vanishes. With increasing distance of atom A from the mirror, the decay rate grows but it is always lower than the respective free-space result (green dot-dashed line) because $\Gamma_{B}^{(1)}= -\Gamma_{B}^{(0)}$, and only the free-space decay term $\Gamma_{A}^{(0)}$ of atom A survives.

\begin{figure}
\begin{centering}
  \includegraphics[scale=0.36]{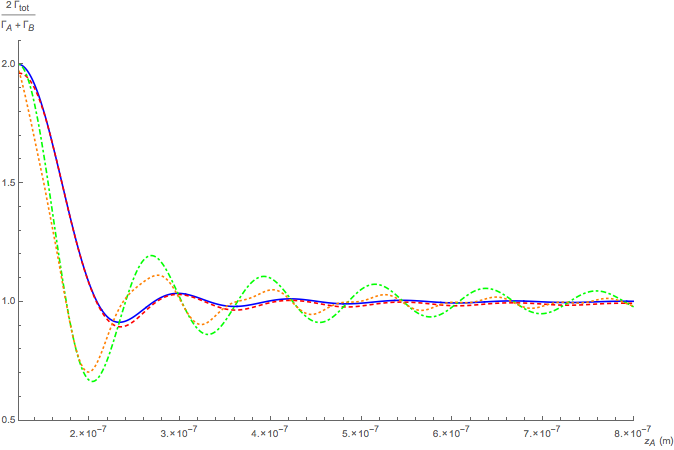}
  \par\end{centering}
 \caption{Scaled collective decay rates (symmetric state) as a function of the position of atom A when atom B is fixed at $z_{B}=1.2\times 10^{-7} \mbox{m}$ (node of the field mode resonant with the atomic transition). The red dashed line and the blue continuous line refer, respectively, to the case when a mirror at $z=0$ is present and free-space case, for dipoles along the $z$ axis. The orange dotted line and the green dot-dashed line are for dipoles along $x$, and are relative to the case when the mirror is present and the free-space case, respectively.}
\label{plotconfrontosimmlambda}
\end{figure}

\begin{figure}
\begin{centering}
  \includegraphics[scale=0.36]{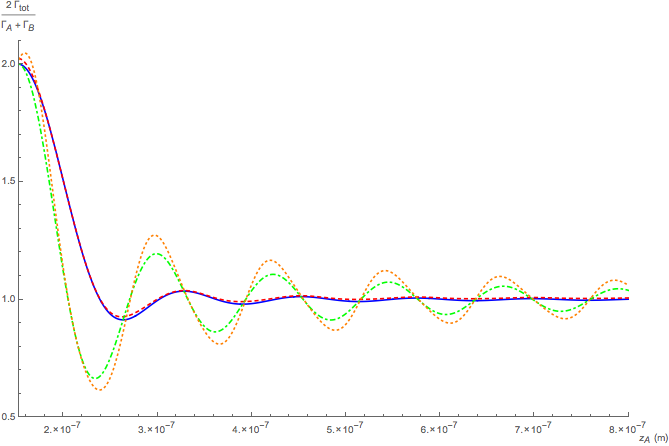}
  \par\end{centering}
  \caption{Scaled collective decay rates (symmetric state) as a function of the position of atom A. Atom B is placed at $z_{B}=1.5\times 10^{-7} \mbox{m}$ (antinode of the field mode resonant with the atomic transition). The red dashed line and the blue continuous line refer, respectively, to the case when a mirror at $z=0$ is present and the free-space case, for dipoles along $z$. The orange dotted line and the green dot-dashed line are for dipoles along $x$, and are relative to the case when the mirror is present and the free-space case, respectively.}
\label{plotconfrontosimmax}
\end{figure}

It is also worth discussing the decay rate for specific positions of the fixed atom B related to the transition wavelength of the two atoms. If we locate the emitter B in a node of the electromagnetic field mode resonant with the atomic transition and vary the position of atom A, we obtain the results shown in Figure \ref{plotconfrontosimmlambda} which should be compared with Figure \ref{dipolizxsimmetrico}. For atomic dipoles oriented along $z$, we observe a better matching between the continuous blue curve (free space) and the red dashed curve (mirror at $z=0$); a similar result is obtained in the case of dipoles along $x$,  given by the green dot-dashed line (free space) and the orange dotted line (mirror at $z=0$). This matching further improves if we locate atom B in an antinode, as shown in Figure \ref{plotconfrontosimmax}. The physical reason for such behaviour might be interpreted in the following way: fixing one atom in a position which is multiple of the atomic transition wavelength $\lambda_{0}$ seems to reduce any eventual positive or negative influence of the boundary on the collective spontaneous emission, if compared to the atomic position effect of Figure \ref{dipolizxsimmetrico}. The mirror has a relevant impact on the free-space oscillatory profile which becomes more regular when atom B is placed in the antinode (comparing the parallel alignment configurations of Figures \ref{plotconfrontosimmlambda} and \ref{plotconfrontosimmax}).

\begin{figure}
\begin{centering}
  \includegraphics[scale=0.36]{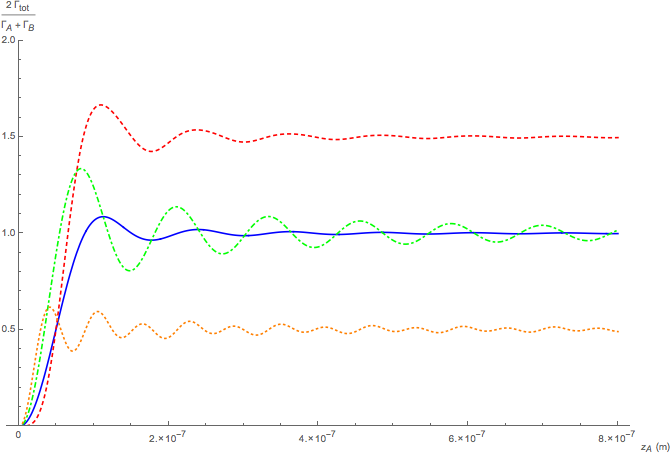}
  \par\end{centering}
  \caption{Scaled collective decay rates (antisymmetric state) as a function of the position of atom A, with atom B fixed at $z_{B}=10 \mathring{A}$. The red dashed line and the blue continuous line refer, respectively, to the case when a mirror at $z=0$ is present and the free-space case, for dipoles along $z$. The orange dotted line and the green dot-dashed line are for dipoles along $x$, and are relative to the case when the mirror is present and the free-space case, respectively.
  }

\label{dipolizxantisimmetrico}
\end{figure}

\begin{figure}
\begin{centering}
  \includegraphics[scale=0.36]{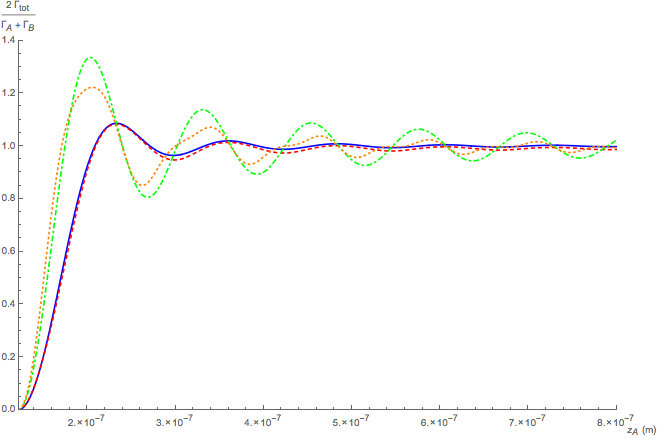}
  \par\end{centering}
  \caption{Scaled collective decay rates (antisymmetric state) as a function of the position of atom A when atom B is fixed at $z_{B}=1.2\times 10^{-7} \mbox{m}$ (node of the field mode resonant with the atomic transition). The red dashed line and the blue continuous line refer, respectively, to the case when a mirror at $z=0$ is present and free-space case, for dipoles along the $z$ axis. The orange dotted line and the green dot-dashed line are for dipoles along $x$, and are relative to the case when the mirror is present and the free-space case, respectively.}

\label{plotconfrontoantisimmlambda}
\end{figure}

\begin{figure}
\begin{centering}
  \includegraphics[scale=0.36]{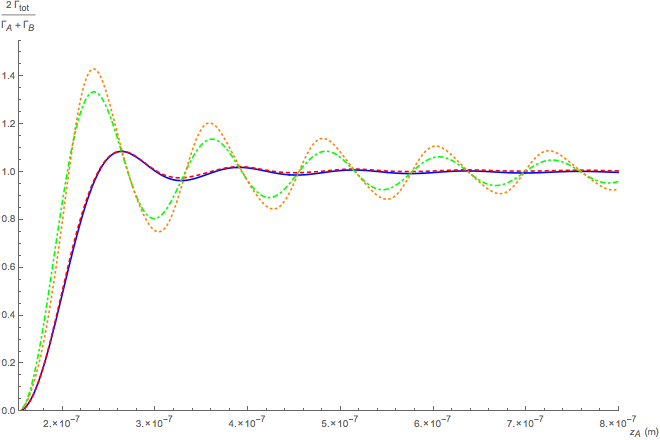}
  \par\end{centering}
   \caption{Scaled collective decay rates (antisymmetric state) as a function of the position of atom A. Atom B is placed at $z_{B}=1.5\times 10^{-7} \mbox{m}$ (antinode of the field mode resonant with the atomic transition). The red dashed line and the blue continuous line refer, respectively, to the case when a mirror at $z=0$ is present and the free-space case, for dipoles along $z$. The orange dotted line and the green dot-dashed line are for dipoles along $x$, and are relative to the case when the mirror is present and the free-space case, respectively.}
\label{plotconfrontoantisimmax}
\end{figure}

Analogous remarks can be made for the plots representing the scaled collective spontaneous emission rate of the two-atom system prepared in the antisymmetric state, as Figures \ref{dipolizxantisimmetrico}, \ref{plotconfrontoantisimmlambda} and \ref{plotconfrontoantisimmax} show.

All this clearly shows that the presence of the mirror can significantly affect the collective radiative behaviour of the two-atom system.

\section{Conclusions}

In this paper we have investigated the spontaneous emission rate of a system composed by two identical entangled atoms in a generic macroscopic environment and interacting with the electromagnetic field. We have considered both symmetric and antisymmetric states of the two emitters.
The first result we have obtained is a general analytical expression of the collective transition rate of our system from the initial symmetric or antisymmetric state to the collective ground state, expressed in terms of the Green's tensor of the electromagnetic field. This expression can be applied to any external environment, whose magnetoelectric properties are contained in the electromagnetic Green's tensor.

We have then considered two specific cases: both atoms in free space and near a perfectly reflecting plate. In free space we have analyzed the decay rate as a function of the interatomic distance, recovering the known results of superradiant and subradiant behaviours. In the non-retarded limit of small interatomic separations, the outcome shows an enhanced emission rate for the symmetric initial state with respect to the case of isolated atoms (superradiance), and the ordinary behaviour of independent atoms for increasing distances. Concerning with the transition rate from the antisymmetric state, it results completely inhibited in the non-retarded limit (subradiance). The enhancement or the inhibition are due to cooperative processes arising from constructive and destructive interference effects between the two correlated atoms, when they are very close to each other.

Subsequently, we have considered the case of two correlated atoms in the presence of a perfectly reflecting planar surface. We have shown that the presence of the boundary significantly affects the superradiant and subradiant decay processes of our atomic system and discussed the results as a function of the interatomic distance as well as of the distance of the atoms from the plate. Our results show that the presence of the plate can enhance or weaken the superradiant/subradiant decay features (if compared to the free-space case) according to the specific orientation of the dipole moments and to atom-plate distances with respect to the atomic transition wavelength. This shows the possibility of controlling and manipulating the collective decay through the external environment.

\ack

R. Palacino wishes to thank G. Baio and J. Hemmerich for stimulating discussions on the subject of the present paper.
This work was supported by the German Research Foundation (DFG, Grants BU 1803/3-1 and GRK 2079/1). S.Y.B is grateful for support by the Freiburg Institute of Advanced Studies.

%
%
%
%
%
%
%
%
%
%
%
%
%
%



\section*{References}
\begin{thebibliography}{99} 
\small

\bibitem[1]{dicke}Dicke R H 1954 \textit{Phys. Rev.} \textbf{93} 99

\bibitem[2]{gross}Gross M and Haroche S 1982 \textit{Phys. Rep.} \textbf{93} 304

\bibitem[3]{peng}Peng J S and Li G X 1998 \textit{Introduction to Modern Quantum Optics} (Singapore: World Scientific Publishing Co. Pte. Ltd.)

\bibitem[4]{persico}Leonardi C, Persico F and Vetri G 1986 \textit{Riv. Nuovo Cimento} \textbf{9} 1

\bibitem[5]{benedict}Benedict M G, Ermolaev A M, Malyshev V A, Sokolov I V and Trifonov E D 1996 \textit{Super-radiance: Multiatomic Coherent Emission} (New York: Taylor Francis Group, LLC)

\bibitem[6]{temnov}Temnov V V and Woggon U 2005 \textit{Phys. Rev. Lett.} \textbf{95} 243602

\bibitem[7]{pan}Pan J, Sandhu S, Huo Y, Stuhrmann N, Povinelli M L, Harris J S, Fejer M M and Fan S 2010 \textit{Phys. Rev. B} \textbf{81} 041101

\bibitem[8]{martin-cano}Mart\'{í}n-Cano D, Mart\'{í}n-Moreno L, Garc\'{í}a-Vidal F J, and Moreno E 2010 \textit{Nano Lett.} \textbf{10} 3129

\bibitem[9]{fleury}Fleury R and Al\'{ù} A 2013 \textit{Phys. Rev. B} \textbf{87} 201101

\bibitem[10]{kastel}K\"{ä}stel J and Fleischauer M, 2005 \textit{Phys. Rev. A} \textbf{71} 011804

\bibitem[11]{choquette}Choquette J J, Marzlin K P and Sanders B C 2010 \textit{Phys. Rev. A} \textbf{82} 023827

\bibitem[12]{casabone}Casabone B, Friebe K, Brandst\"{ä}tter B, Sch\"{ü}ppert K, Blatt R and Northup T E 2015 \textit{Phys. Rev. Lett.} \textbf{114} 023602

\bibitem[13]{chaneliere}Chaneliére T, Matsukevich D N, Jenkins S D, Kennedy T A B, Chapman M S and Kuzmich A 2006 \textit{Phys. Rev. Lett.} \textbf{96} 093604

\bibitem[14]{jen}Jen H H 2012 \textit{Phys. Rev. A} \textbf{85} 013835

\bibitem[15]{ralf}R\"{o}hlsberger R, Schlage K, Sahoo B, Couet S and R\"{u}ffer R 2010 \textit{Science} \textbf{328} 1248

\bibitem[16]{keaveney}Keaveney J, Sargsyan A, Krohn U, Hughes I G, Sarkisyan D and Adams C S 2012 \textit{Phys. Rev. Lett.} \textbf{108} 173601

\bibitem[17]{pellegrino}Pellegrino J, Bourgain R, Jennewein S, Sortais Y R P, Browaeys A, Jenkins S D and Ruostekoski J 2014 \textit{Phys. Rev. Lett.} \textbf{113} 133602

\bibitem[18]{crubellier}Pavolini D, Crubellier A, Pillet P, Cabaret L and Liberman S 1985 \textit{Phys. Rev. Lett.}
\textbf{54} 1917

\bibitem[19]{devoe}DeVoe R G and Brewer R G 1996 \textit{Phys. Rev. Lett.} \textbf{76} 2049

\bibitem[20]{petrosyan}Petrosyan D, Kurizki G 2002 \textit{Phys. Rev. Lett.} \textbf{89} 207902

\bibitem[21]{takasu}Takasu Y, Saito Y, Takahashi Y, Borkowski M, Ciurylo R and Julienne P S 2012 \textit{Phys. Rev. Lett.} \textbf{108} 173002

\bibitem[22]{filipp}Filipp S, Van Loo A F, Baur M, Steffen L and Wallraff A 2011 \textit{Phys. Rev. A} \textbf{84} 061805

\bibitem[23]{buhmann}Buhmann S Y 2012 \textit{Dispersion Forces I} (Springer, Vol 247)

\bibitem[24]{knoll}Kn\"{ö}ll L, Scheel S and Welsh D G 2001 in: \textit{Coherence and Statistics of Photons and Atoms} (New York: Wiley)

\bibitem[25]{scheel}Scheel S and Buhmann S Y 2008 \textit{Acta Physica Slovaca} \textbf{58} 675

\bibitem[26]{santra}Santra R and Cederbaum L S 2002 \textit{Phys. Rep.} \textbf{368} 10

\bibitem[27]{pablo}Barcellona P, Passante R, Rizzuto L and Buhmann S Y 2016 \textit{Phys. Rev. A} \textbf{93} 032508

\bibitem[28]{pablo1}Barcellona P, Passante R, Rizzuto L and Buhmann S Y 2016 \textit{Phys. Rev. A} \textbf{94} 012705

\bibitem[29]{haakh}Haakh H R, Henkel C, Spagnolo S, Rizzuto L and Passante R 2014 \textit{Phys. Rev. A} \textbf{89} 022509

\bibitem[30]{tanas}Tana\'{s} R and Ficek Z 2004 \textit{Journ. Opt. B} \textbf{6} 2

\bibitem[31]{arias}Arias E, Duenas J G, Menezes G and Svaiter N F 2016 \textit{Journ. of High Energy Phys.} \textbf{5} 1


\bibitem[32]{skribanowitz}Skribanowitz N, Herman I P , MacGillivray J C and Feld M S 1973 \textit{Phys. Rev. Lett.} \textbf{30} 309

\bibitem[33]{Scheibner}Scheibner M, Schmidt T, Worschech L, Forchel A, Bacher G, Passow T and Hommel D 2007 {\it Nature Phys.} {\bf 3} 106

\bibitem[34]{Novotny11}Novotny L and van Hulst N 2011 \textit{Nature Photon.} \textbf{5} 83

\end{thebibliography}
\end{document}